\documentclass[preprint2]{aastex}

\def\dsec{{\rlap.}$^{\prime\prime}$}

\shorttitle{ULXs in NGC~3256}
\shortauthors{Neff, Ulvestad, \& Campion}
\journalid{}{}
\articleid{}{}
\def\H2{\ion{H}{2}}

\begin{document}


\title{Radio Emission Associated with Ultraluminous 
         X-ray Sources in the Galaxy Merger NGC~3256 }

\author{S. G. Neff } 
\affil{Laboratory for Astronomy and Solar Physics, NASA's Goddard Space Flight 
Center, Greenbelt, MD 20771}
\email{susan.g.neff@nasa.gov}

\author{James S. Ulvestad}
\affil{National Radio Astronomy Observatory, P. O. Box 0, Socorro, NM 87801} 
\email{julvesta@aoc.nrao.edu}

\and

\author{S. D. Campion }
\affil{SSAI and GSFC }
\email{scampion@hst.nasa.gov}




\clearpage

\begin{abstract}
We present new 6, 3.6, and 2~cm VLA radio observations of 
the nearby merger 
system NGC~3256, with resolutions of $\sim 100$~pc, which
reveal compact radio sources 
embedded in more diffuse emission at all three wavelengths. 
The two radio nuclei are partially
resolved, but the two dominant compact sources that remain 
coincide with the two most powerful
compact Ultraluminous X-ray sources (ULXs) recently
reported by Lira {\it et al.}  The radio/X-ray ratios for
these two sources are too high by factors of $>$100--1000
to be normal X-ray binaries.  However, their radio and 
X-ray powers and ratios are consistent with  
low-luminosity active galactic nuclei (LLAGNs), and optical
emission lines suggest the presence of a nuclear disk around
the northern nucleus.  If the two nuclear ULXs 
are LLAGNs, their associated black holes are separated by only
$\sim$1kpc, about 6 times closer to one another than those 
found recently in the merger galaxy NGC~6240.  
A third ULX on the outskirts of the merger is also a radio
source, and probably is a collection of supernova remnants.
The remaining ULXs are not coincident with any
source of compact radio emission, and are
consistent with expectations for beamed X-ray binaries or 
intermediate-mass black holes.

\end{abstract}
\keywords{galaxies: individual (NGC 3256) --- galaxies: interactions --- 
galaxies: starburst --- galaxies: star clusters --- X-rays: binaries --- X-rays: 
galaxies}

\section{Introduction}

   NGC~3256 is a well-studied merging galaxy system and is the
brightest IR source in the nearby universe, with a
luminosity of $3\times 10^{11} L_\odot$ in the 8--1000~$\mu$m
range \citep{Sar89}, after conversion to a distance of 37~Mpc 
(H$_0$ = 75 km s$^{-1}$ Mpc$^{-1}$).  
It is also at the top end of the X-ray luminosity range 
for starburst galaxies \citep{Lir02}. 
Tidal tails extending $\sim$ 200 kpc and other morphological 
and kinematical evidence \citep{deV61,Too77,Fea78,Sch86} 
suggest that the system was composed of two gas-rich disks, 
which coalesced $\sim$1 galaxy crossing time ago \citep{Kee95}.
However, two ``nuclei'' are detected at radio and NIR wavelengths,
suggesting that the progenitors may not have completed their 
merger \citep{Nor95,Kot96}.   
All observations to date suggest that the system properties are 
best explained by an ongoing very intense starburst and associated
superwind \citep{Lip00}.

   Recently, high-resolution {\it Chandra} observations of 
NGC~3256 detected 14 discrete hard X-ray sources with luminosities
of 10$^{39}$--10$^{40.5}$ ergs s$^{-1}$ at an assumed distance
of 56~Mpc \citep{Lir02}; 13 of these remain above $10^{39}$~ergs~s$^{-1}$
for a distance of 37~Mpc.  Such sources are referred to as
Ultraluminous X-ray sources (ULXs), since their isotropic
luminosities are far above the Eddington limit of 
$2\times 10^{38}$~ergs~s$^{-1}$ for an accreting $1.4M_\odot$
neutron star in an X-ray binary, but below the limit for classical active
galactic nuclei (AGNs), $\sim 10^{42}$~ergs~s$^{-1}$.
Possible explanations for ULXs include 
accreting intermediate-mass ($\sim$100--1000 M$_{\odot}$) black holes (IMBH;
probably in binaries), beamed ``normal'' X-ray binaries, microquasars, 
transient super-Eddington accretors, young supernova 
remnants (SNRs) in extremely 
dense environments, hypernovae/GRB's, or background BL Lac objects
\citep{Fab89,Mar95,Col99,Mak00,Kin01,Kin02,Rob03}.  The first 
large samples of ULXs 
suggested that they are fairly rare in normal galaxies 
\citep{Rob00}, but much more common, and brighter, in merging and starburst 
galaxies \citep{Fab01,Kaa01}.  
Efforts  to identify ULX counterparts at other wavelengths
have had limited success:  a ULX in NGC~5204 may be 
associated with a young optical star cluster
\citep{Rob01,Goa02}, at least four ULXs are  
associated with X-ray ionized nebulae or SNR  \citep{Pak02, Rob03},
several ULXs in elliptical galaxies may be in 
globular clusters \citep{Wu02, Ang01}, 
and a radio counterpart recently has been identified for a 
ULX in NGC~5408 \citep{Kaa03}.

  Centimetric radio observations penetrate the dust
causing the optical obscuration in starbursts. 
If radio emission is associated with a ULX, 
the radio luminosity and spectrum can be used to constrain 
possibilities for the ULX identity.
Previous radio imaging of NGC~3256 identified two bright steep-spectrum
nuclear sources \citep{Nor95}, but had insufficient resolution 
to probe the ULX environment.
Here, we present new, high-resolution radio images of NGC~3256
that provide new clues to the nature of its ULXs.

\section{Radio Images and Measurements}

 NGC~3256 was observed in 2001 February ({\bf BnA} configuration)
and 2002 April ({\bf A} configuration) using the 
NRAO\footnote{The National Radio
Astronomy Observatory is a facility of the National Science
Foundation operated under cooperative agreement by Associated
Universities, Inc.} Very Large Array (VLA) at 6~cm (4860~MHz) and 
3.6~cm (8460~MHz), for $\sim$ 7.5hrs each. 
A 12-minute 2~cm (14940~MHz) observation was obtained in 
2002 April, and a 35-minute 2~cm observation at lower
resolution ({\bf CnB} configuration) from 1986 was retrieved from 
the VLA archive.
The data were calibrated, self-calibrated, and imaged as 
described in \citet{Nef00}.   Typical resolutions (FWHM)
are 50-60~pc ($\sim$0\farcs35) by 140--180~pc (0\farcs7--1\farcs0).

We measured flux densities of compact radio sources 
associated with ULXs by fitting a parabola to the peak,
and derived the amount of diffuse
emission around the selected sources by integrating
over a specified box and subtracting the flux of
the related compact source (Table~\ref{tab:radio}). 
Errors in radio positions are $\lesssim$0\farcs1, 
negligible compared to the X-ray position 
uncertainties\footnote{see the {\it Chandra} Proposer's Guide 
v 4.0, 2001, http://asc.harvard.edu/udocs/docs/POG/MPOG} 
of $\gtrsim$0\farcs6;
radio flux densities have typical ($1\sigma$) errors 
of 5\% at 3.6 and 6~cm, and 10\% at 2~cm.

At both 6~cm and 3.6~cm, two bright compact sources are 
embedded in diffuse extended emission (Figure 1). No radio emission is 
detected outside the central 40\arcsec\ (7~kpc) of the galaxy,
with respective $3\sigma$ limits of $\sim$36 and $\sim$30~$\mu$Jy 
beam$^{-1}$ at 6 and 3.6~cm.   At the highest
resolution (Figure 2, {\bf A} configuration only),  
3.6 and 2~cm images resolve both of
the bright nuclear sources into several discrete 
components.  ULX positions from \citet{Lir02} are indicated on each 
image (Figures~1 and 2), with the \citet{Lir02} source numbers  
used in the remainder of this paper.  

\section{Identification of Radio Sources Associated with ULXs}
ULXs 7 and 8 are coincident with compact radio sources, and ULX~13 
is at the location of an extended, partly resolved radio source.
Properties of these radio sources are given in Table~\ref{tab:radio}.
Radio emission is found at the edges of the 1\farcs8 radius ($3\sigma$) 
{\it Chandra} error circles of several other ULXs in NGC~3256, 
but is neither compact nor precisely coincident with the X-ray position 
(see Fig.~1): for example, ULXs 9, 10, and 11 are clustered
around the edges of a group of compact radio sources, 
ULX~2 is near the edge of another, as shown in Figure 1. 
\citet{Lir02} note that these are sites of massive
 stars ionizing their surroundings, as 
evidenced by strong H$\alpha$ emission \citep{Lip00}.  
The nominal separations of $\sim$ 2\arcsec\ between radio
emission and ULXs correspond to $\sim 360$ pc in NGC 3256.
 
Figure~2 shows that ULXs 7 and 8 nominally 
coincide with the two most compact nuclear radio peaks. 
\citet{Lir02} indicate that the northern X-ray 
source, ULX~7, is resolved, with a size of 
$\sim$1\arcsec--1\farcs5 ($\sim$180--270 pc); they 
suggest that this corresponds to ``a compact grouping of individual 
sources $\ldots$ at the heart of the nuclear starburst." However,
Mushotzsky (priv. comm) has determined that ULX~7 is a dominant
point source (unresolved by {\it Chandra}) embedded in low-level extended 
emission,
as can be seen in Fig 2c.
Our 3.6 and 2~cm images (Fig.~2a,b) indicate that the radio emission 
at this position is partially resolved on a similar scale 
and is also embedded in extended emission.

\section{Nature of the Ultraluminous X-ray Sources}

    The radio sources near ULXs 7 and 8 have respective spectral 
indices of $\alpha=-1.04$ and $\alpha=-0.74$ (for 
$S_{\nu}\propto\nu^{\alpha}$) between 3.6 and 2~cm, 
indicative of optically thin 
synchrotron emission.  However, the source spectra flatten between
6 and 3.6~cm in matched-resolution images 
($\alpha=-0.49$ and $\alpha=-0.16$), which may indicate that some
of the subcomponents are free-free absorbed at longer 
wavelengths.  For gas at 6000~K \citep{Lip00}, the required
emission measure for significant free-free absorption at
5~GHz is about $5\times 10^7$~cm$^{-6}$~pc. This implies (for example)
a typical ionized density of $\sim 7000$~cm$^{-3}$ for a path length
of $\sim 1$~pc, or $\sim 700$~cm$^{-3}$ for a path length of 100~pc.
The density in the latter case is similar to that found by 
\citet{Lip00} on scales of a few hundred parsecs, but the relatively
low filling factor of the gas seen on these large scales makes it
likely that any free-free absorption would be caused by higher density
gas on scales of a parsec or less.
 
We use the ratio of 6-cm radio emission to 2--10~keV X-ray emission,  
$R_X\equiv L_R/L_X = \nu L_\nu({\rm 5\ GHz})/L_X(2-10\ {\rm keV})$, 
defined by \citet{Ter03}, to distinguish further among possible 
origins of ULXs 7 and 8.  Table~\ref{tab:rxray} gives values of this ratio
for ULXs in NGC~3256 and for a number of comparison objects.
First, consider flaring X-ray transients.  ULXs 7 and 8 might be either
beamed ``normal'' binaries or unbeamed binaries including
intermediate-mass black holes ($\sim 100$--$1000M_\odot$).
Using the peak values in both wavebands, \citet{Fen01} compile values of 
the radio/X-ray ratio for galactic X-ray transients in outburst, and 
show that it always is less than $10^4$~mJy~Crab$^{-1}$, corresponding
to $R_X < 2.3\times 10^{-5}$.  This is consistent with the result of 
\citet{Kaa03},
who interpret a ULX in NGC~5408 as a beamed X-ray transient (although
they note it could also be an IMBH); the faint
radio detection implies $R_X < 10^{-5}$.  In contrast,
ULXs 7 and 8 in NGC~3256 have 
$R_X$ well above $10^{-3}$, much larger than 
the ratio for galactic X-ray transients; the radio flux densities are 
more than $10^5$ times stronger than expected from Cygnus X-1
beamed toward us from a distance of 37~Mpc \citep{Geo02}.

The northern nucleus of NGC~3256 (ULX~7) is known to harbor an
extreme starburst \citep{Lip00}, so it is worth considering
whether the radio and X-ray emission can be produced by 
only young stars or associated high mass X-ray binaries (HMXB)
Optical and UV emission lines suggest that the ensemble of stars in
the northern nucleus has $T_{\rm eff}=35,000$ K \citep{Lip00}, 
and the X-ray spectrum is consistent
with a hot-gas component at $\sim 2\times10^7$K \citep{Lir02}.
A strong upper limit to the number of stars present
can be calculated by assuming that the entire 3~mJy flux density at
2~cm is due to thermal emission from ionized gas (unlikely due to 
the steep radio spectrum); this would require the equivalent
of $\sim 10^5$~O7 stars \citep{Vac94}.  These hot stars would 
generate X-ray emission primarily in stellar winds;
the expected X-ray luminosity from
$10^5$ O7 stars would be $\sim 10^{37}$~ergs~s$^{-1}$ 
\citep{Lon80,Chl89}, a factor of 1000 below the observed
X-ray luminosity.  It is likely that a cluster of O stars will
contain a significant number of HMXB's.  If systems 
like Cyg X-1 \citep{Geo02} were to produce all of the 
ULX~7 X-ray emission, $\sim$600-1000 such systems would be required,
1 per 100 O stars.  These HMXBs would produce $<10^{-4}$
of the observed radio emission.
Although it is possible that the radio and X-ray emission
could come from a combination of O stars and many HMXBs, it seems
unlikely that this many HMXB's would have already formed while
there are still $\sim$10$^{5}$ unevolved O stars. 
 
  Since the X-ray and radio sources in ULX 7 both contain multiple
components (diffuse and discrete), it seems useful to compare $R_X$ 
for ULX 7 with better-resolved starburst region in a nearby galaxy. 
NGC~253, at a distance of $\sim$ 2.6~Mpc, has an
X-ray flux of $\sim 2.5\times 10^{-12}$ ergs~s$^{-1}$ (Weaver
et al. 2002) in its central 5\arcsec\ (60 pc); integrating over
the same area in the 5~GHz VLA image from Antonucci \& Ulvestad
(1988) yields a flux density of $\sim 600$~mJy.  For this compact
nuclear starburst, therefore, $R_X\sim 1\times 10^{-2}$, a factor
of only a few higher than ULX 7.  This suggests that ULX 7 
may represent a region of extreme ``nuclear'' star formation.  

   Next, we consider the possibility that ULXs 7 and 8
produce X-ray and radio emission in multiple 
SNRs.  For the galactic SNR Cas~A, 
$L_R/L_X\approx 0.02$, similar to the values for ULXs
7 and 8 (Cas-A X-ray flux from 
http://hea-www.harvard.edu/ChandraSNR /G111.7-02.1/; contemporaneous 
radio flux density derived from Baars et al. 1977).  The total 
radio and X-ray fluxes are $\sim 1000$ times stronger than Cas~A 
would be at 37~Mpc, implying that many similar SNRs would be 
required within a volume of $10^5$--$10^6$~pc$^3$.  In the case of 
ULX 13, associated with a partially resolved steep-spectrum radio source 
well away from the galaxy nuclei, $R_X \sim 0.01$, also consistent with 
Cas~A, with total emission $\sim 200$ times more powerful.  Some young radio 
supernovae may have radio powers of tens to hundreds times that of Cas~A
\citep{Wei86}, but accounting for the relatively steady 
radio emission from the NGC~3256 nuclei still would require 
many supernovae over at least the last 15~yr.  The same arguments
apply to the case of hypernovae \citep{Rob03}, with the additional
constraint that NGC~3256 has not been identified with any GRB's.

A final possibility is that ULXs 7 and 8 are low-luminosity 
AGNs (LLAGNs), generated in or fueled by the galaxy 
merger.  \citet{Lir02} note that the X-ray properties and overall
SED of ULX 8 are consistent with this possibility, but 
suggest that ULX~7 is to be too strong at 10$\mu$m to be an AGN.
However, \citet{Ho99} shows that
LLAGNs have a maximum in the SED at $\sim$10$\mu$m.
\citet{Ho99} also finds, when different classes 
of AGN SEDs are normalized at 1 keV, that
LLAGNs are considerably stronger at mid-IR wavelengths than
other classes of AGN (\citet{Lir02} made the SED comparison 
by normalizing various AGN SEDs at 3.5$\mu$m).   
Thus, \citet{Ho99}'s work stongly supports the idea that
ULX~7 is also a LLAGN. Both ULXs~7 and 8 have values of 
$R_X$ well within the range for radio-loud
AGNs, defined by \citet{Ter03}.  $R_X$ for ULXs 7 and 8 is a factor 
of a few larger than for several LLAGNs at comparable distances 
\citep{Ulv01} and for the nearby galaxy nucleus in M32 \citep{Ho03a}, 
and is similar to the values for the two AGNs in the merger NGC~6240 
\citep{Kom03}.  Further evidence for the LLAGN possibility comes
from examination of archival HST long-slit spectra of the northern 
nucleus, which show broadened H$\alpha$, H$\beta$, [N II], and
[S II] emission lines with linewidths of $\sim$450 km sec$^{-1}$,
as shown in Figure 3.  The emission line  
show  a rotation-curve-like shear in their
peak velocities of $\sim$300km sec$^{-1}$ over a projected distance
of $\sim$40pc.  If we assume this shear indicates rotation of a
nuclear disk, it implies that ULX 7(N) has a mass of {\it at least} 
10$^{8}$M$_{\odot}$ within $\sim 40$ pc.

LLAGNs are the most plausible explanation for ULXs~7 and 8 
in NGC~3256, although large collections of supernova remnants
are also consistent with their radio/X-ray ratios and powers. 
If ULXs~7 and 8 are LLAGN, the dual black holes
are separated by only $\sim 1$~kpc, considerably closer than the
$\sim 6$~kpc separation for the dual AGNs in NGC~6240.

For the remaining ULXs that are not associated with
discrete radio sources, the radio/X-ray limits are consistent with
expectations for high-mass X-ray binaries, either beamed or
accreting at super-Eddington rates, and ULX 13 is also consistent
with a collection of confined SNRs. 
 
\section{Summary}

We have presented new, high resolution, sensitive radio images of
the merging galaxy NGC~3256.  We found that three compact radio
sources appear to be identified with ULX's.  Two of the ULX's
are coincident with the two galaxy nuclei, and have properties
indicative of being two LLAGN, separated by $\sim 6$ pc. 
Radio properties of a third ULX suggest that is is a
large complex of supernova remnants, and the remaining
ULX's are likely associated with HMXRBs or IMBHs.


\acknowledgments

Part of this work was supported by NASA grant 344-01-21-77.  
We thank Vivek Dhawan, Phillip Kaaret, and Richard Mushotzky 
for useful discussions.  SGN and JSU acknowledge the Aspen
Center for Physics for hospitality during the preparation of this work.
This research has made use of the NASA/IPAC Extragalactic Database 
(operated by the Jet Propulsion Laboratory,
California Institute of Technology, under contract with the 
National Aeronautics and Space Administration) and of NASA's
Astrophysics Data System Bibliographic Services.

\clearpage

%
%
%

\begin{deluxetable}{cccccccl}
%
%
%
\tablewidth{0pt}
\tablecaption{Discrete Radio Sources at ULX Positions}
\tablehead{
\colhead{ULX }  &
\colhead{R. A. }  &    
\colhead{Decl. }  &  
\colhead{S$_{\rm 6cm}$ } &     
\colhead{S$_{\rm 3.6cm}$ } & 
\colhead{S$_{\rm 2cm}$ } &
\colhead{Resolution }  \\
\colhead{}&\colhead{(J2000)}&\colhead{(J2000)}&
\colhead{(mJy)}& \colhead{(mJy)}& \colhead{(mJy)}&
\colhead{(Type)} &  
} 
\startdata
7(N) & 10 27 51.23 & -43 54 14.0 & \nodata & $2.3 \pm .1$ & $1.4 \pm .2$ &  
0\farcs63 $\times$ 0\farcs15 ~(peak) \\

 &  &  & $5.0 \pm .3 $ & $3.8 \pm .2$ & $2.1\pm .2$ &
1\farcs00 $\times$ 0\farcs26 ~(peak) \\

 &  &  & $31 \pm 2 $ & $20 \pm 1 $ & $17 \pm 2 $ & 
$\sim$ 2\farcs5 $\times$ ~3\farcs5 ~(box)\\

8(S) & 10 27 51.22 & -43 54 19.2 & \nodata & $3.9 \pm 0.2$ & $3.0 \pm 0.3$ & 
0\farcs63 $\times$ 0\farcs15 ~(peak)\\

 & & & $  6.0 \pm 0.3$ & $5.5 \pm 0.3$ & $3.6 \pm 0.4$ & 
1\farcs00 $\times$ 0\farcs26 ~(peak)\\

 & & & $ 25 \pm 1 $ & $16.1 \pm 0.8 $ & $11 \pm 1 $ & 
$\sim$ 2\farcs5 $\times$ ~3\farcs5 ~(box)\\

13 & 10 27 52.91 & -43 54 11.5 & \nodata & $<0.08 $ & $<0.24 $ & 
0\farcs63 $\times$ 0\farcs15 ~(peak)\\

 & & &  $0.28 \pm 0.03 $ & $0.18 \pm 0.02$ & $<0.36 $ & 
1\farcs00 $\times$ 0\farcs26 ~(peak)\\

 & & &  $1.35 \pm 0.07 $ & $0.88 \pm 0.05$ & $0.5 \pm 0.2 $ & 
$\sim$ 2\farcs5 $\times$ ~3\farcs5 ~(box)\\

\enddata
\tablenotetext{a}{Peak flux is measured using two different 
beam sizes (resolution is FWHM) so that spectral indices may 
be determined for both $\alpha_{3.6}^{6}$ and for $\alpha_{2}^{3.6}$.
The smaller beam size is for A-configuration images with 
restoring beam size (FWHM) 0\farcs63 $\times$  0\farcs15, 
which cannot be obtained at 6~cm.  The larger beam
size given is the smallest possible at 6~cm. 
Integrated flux measures, within boxes, are also 
given; they are used determine diffuse emission spectral indices.}
\label{tab:radio}
\end{deluxetable}


\begin{deluxetable}{lcccccc}
%
%
\tablewidth{0pt}
\tablecaption{Radio/X-ray Properties of ULXs and Comparison Objects}
\tablehead{
\colhead{Source }  &         
\colhead{$\alpha_{\rm 6-3.6}$ }  &
\colhead{$\alpha_{\rm 3.6-2}$ }  &          
\colhead{$S({\rm 5\ GHz})$\tablenotemark{a}  }  &
\colhead{$F_{\rm X}$(2--10 keV)}      & 
\colhead{$R_X$\tablenotemark{b} } &
\colhead{References\tablenotemark{c} } \\
\colhead{}&\colhead{}&\colhead{}&\colhead{(mJy)}&
\colhead{(ergs cm$^{-2}$ s$^{-1}$)}&\colhead{}&\colhead{}   
} 
\startdata 
ULX 7(N) compact & $-0.49$ & $-1.04$ & 5.0 & $1\times 10^{-13}$ & 
$2\times10^{-3}$ & 1,2\\
ULX 8(S) compact & $-0.16$ & $-0.74$ & 6.0 & $3\times 10^{-14}$ & 
$1\times10^{-2}$ & 1,2    \\	  
ULX 7(N) diffuse & $-0.84$ & $-0.18$ & 26 & \nodata & \nodata & 1 \\
ULX 8(S) diffuse & $-1.01$ & $-0.68$ & 19 & \nodata   & \nodata  &  1 \\
ULX 13 diffuse   & $-0.77$ & $-2.03$ & 1.1 & $5\times 10^{-15}$ & $1\times 
10^{-2}$ & 
1,2 \\
Other NGC 3256 ULXs &\nodata  & \nodata & $< 0.05$& $\sim 10^{-14}$ & 
$< 2\times 10^{-3}$ & 1,2 \\
 & & & & & \\
X-ray transients & \nodata & \nodata & \nodata & \nodata &  $ < 2\times 
10^{-5}$ 
 &  
3 \\
NGC~5408 ULX &\nodata & \nodata & 0.3 & $3\times10^{-12}$ & 
$5\times10^{-6}$ & 4 \\
$10^5$ O stars at 37 Mpc & $\sim -0.1$ & $\sim -0.1$ & 
$\sim 3$ & $6\times 10^{-17}$  & $\sim$ 3  & 1,5,6 \\
Cas A  & $\sim -0.5$ & $\sim -0.5$ &  $7\times 10^5$ & $2\times10^{-9}$ & 
$2\times 10^{-2}$  & 7,8 \\
NGC~253 (nuc) & $\sim$-0.7 & ... & 600 & $2\times 10^{-13}$ & 
$1 \times10^{-2}$ & 9,10 \\
LLAGNs at 20 Mpc &  $\sim +0.3$ & $\sim +0.3$ & $\sim 10$ & 
$\sim 1\times 10^{-14}$ & $\sim 1\times 10^{-3}$ & 11 \\
NGC~6240N & $-0.6$ & $-0.6$ & 16 & $8\times10^{-13}$ & $1\times 10^{-3}$ & 
12,13 
\\
NGC~6240S & $-0.7$ & $-0.7$ & 37  & $3\times10^{-13}$ & $6\times 10^{-3}$ & 
12,13 
\\

\enddata
\tablenotetext{a}{Compact radio flux densities measured in a beam
1\farcs00 $\times$ 0\farcs26 (from Table~\ref{tab:radio}).} 
\tablenotetext{b}{$R_X=\nu L_\nu({\rm 5\ GHz})/L_X$, as defined by
\citet{Ter03}.}
\tablenotetext{c}{References: 1. This paper. 2. \citet{Lir02}. 
                 3. \citet{Fen01}.
                 4. \citet{Kaa03}. 
                 5. \citet{Lon80}.
                 6. Chlebowski et al. (1989).
                 7. \citet{Baa77}. 
                 8. http://hea-www.harvard.edu/ChandraSNR/G111.7-02.1/.
                 9.  \citet{Ant88}
                 10. \citet{Wea02}
                 11. \citet{Ulv01}.
                 12. \citet{Bes01}.
                 13. \citet{Kom03}. }
\label{tab:rxray}
\end{deluxetable}

\clearpage



\begin{figure}
\plotone{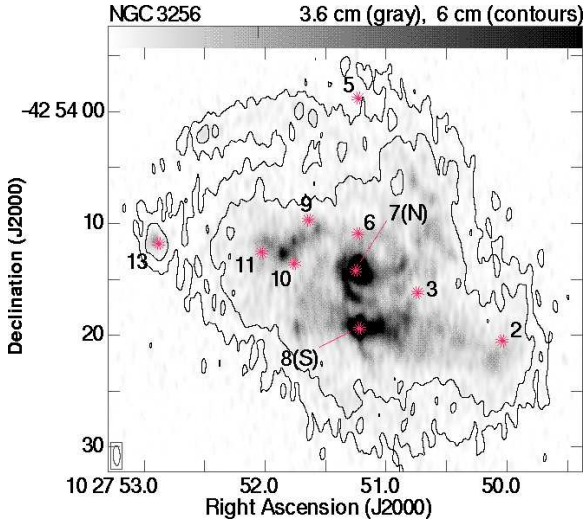}

\caption{ 6~cm (contours) and 3.6~cm (grayscale) image of NGC~3256,
showing the location of the radio emission relative to the ULX's. 
Stars show positions of ULXs detected
by \citet{Lir02}, with 0\farcs6 radius indicating the $1\sigma$
{\it Chandra} position uncertainty.
The restoring beams are 1\farcs78 $\times$ 0\farcs60 (6~cm) and
1\farcs01 $\times$ 0\farcs39 (3.6~cm), with rms $\sim$13$\mu$Jy beam$^{-1}$
and $\sim$9$\mu$Jy beam$^{-1}$ respectively.  
Contours are at 0.07 and 0.15mJy beam$^{-1}$, gray scale  
shown ranges from 0.03 to 0.60mJy beam$^{-1}$, peak flux
in 6~cm image is 10.86mJy beam$^{-1}$, peak flux in 3.6~cm image
is 6.18mJy beam$^{-1}$.  }

\label{fig1}
\end{figure}


\begin{figure}
\epsscale{0.50}
\plotone{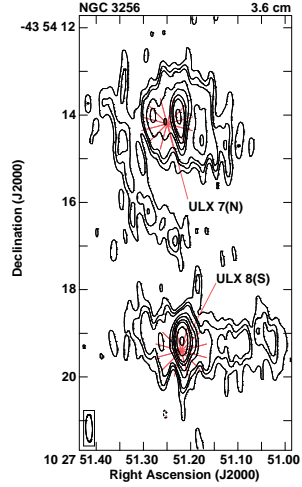}

\caption{3.6~cm, 2.0cm, and X-ray (1.5-10keV) images of the nuclear 
region of NGC~3256.   Stars mark the positions of ULXs 7
(north nucleus) and 8 (south nucleus) 
from \citet{Lir02}, with the 0\farcs6 radius 
corresponding to the $1\sigma$ {\it Chandra} position
uncertainty.   Radio images use only data from April 2002, to provide
the highest resolution. (a) The 3.6~cm image was made with 
uniform weighting (rms $\sim$27$\mu$Jy beam$^{-1}$) and has a
restoring beam of 0\farcs63 $\times$ 0\farcs15; contour levels 
are -0.10, 0.10, 0.15, 0.20, 0.40, 0.60, 0.90, 1.20, 2.00, and 3.500 
mJy beam$^{-1}$.  (b) The 2.0~cm image was made with 12 minutes of
data and uses natural weighting (rms $\sim$85$\mu$Jy beam$^{-1}$); 
contour levels are -0.25, 0.25, 0.40, 0.70, 1.00, 1.50, 2.50 
mJy beam$^{-1}$.  (c) The 1.5-10keV image was made from 28ks of
Chandra data, retrieved from the {\it Chandra} public archive; 
contours are at 6, 8, 10, 12, 15, 18, 24, 30, 45, 60, 75 counts. }

\label{fig2a}
\end{figure}

\setcounter{figure}{1}
\begin{figure}
\epsscale{0.50}
\plotone{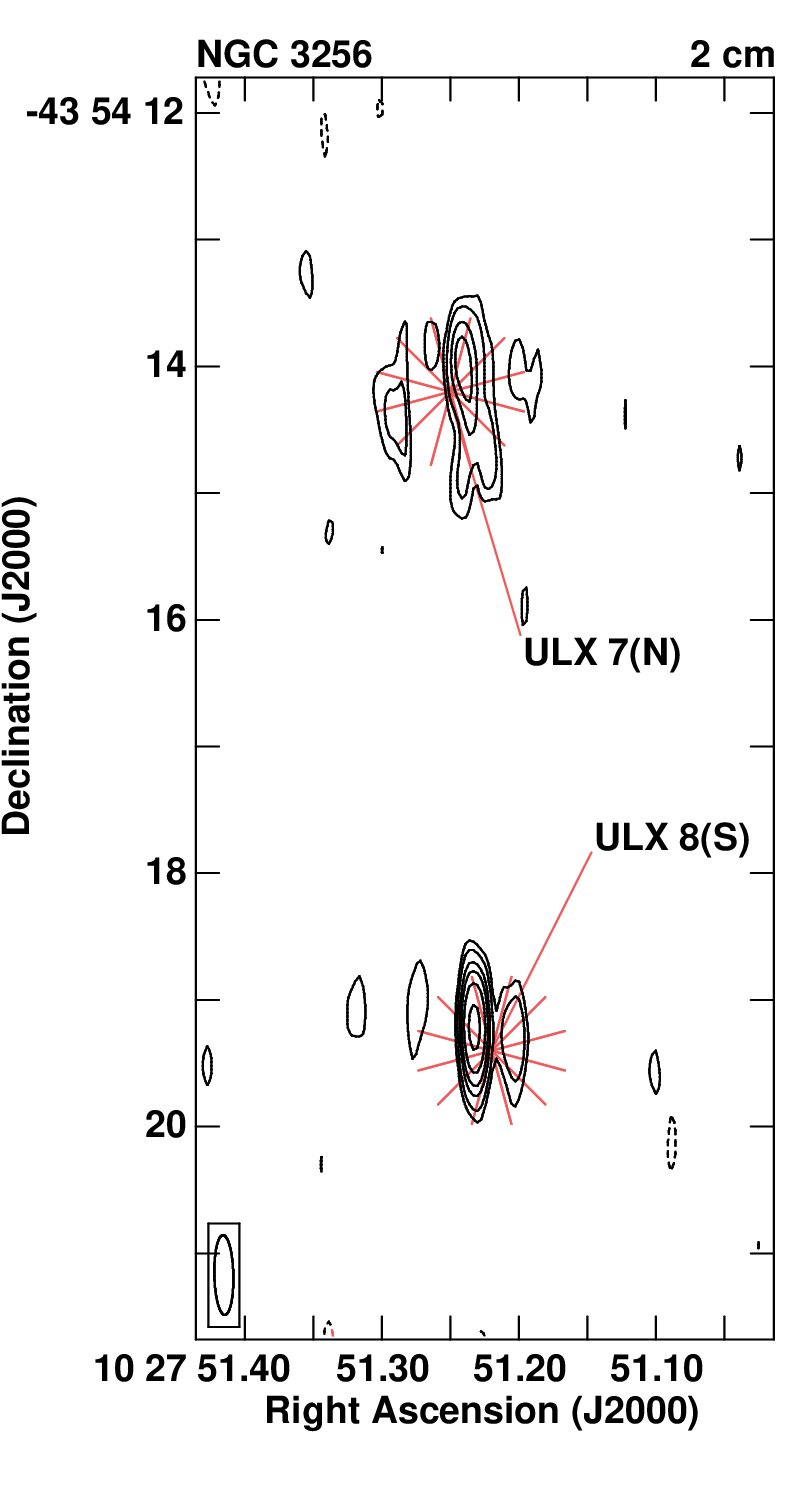}
\caption{Continued}
\label{fig2b}
\end{figure}

\setcounter{figure}{1}

\begin{figure}
\epsscale{0.50}
\plotone{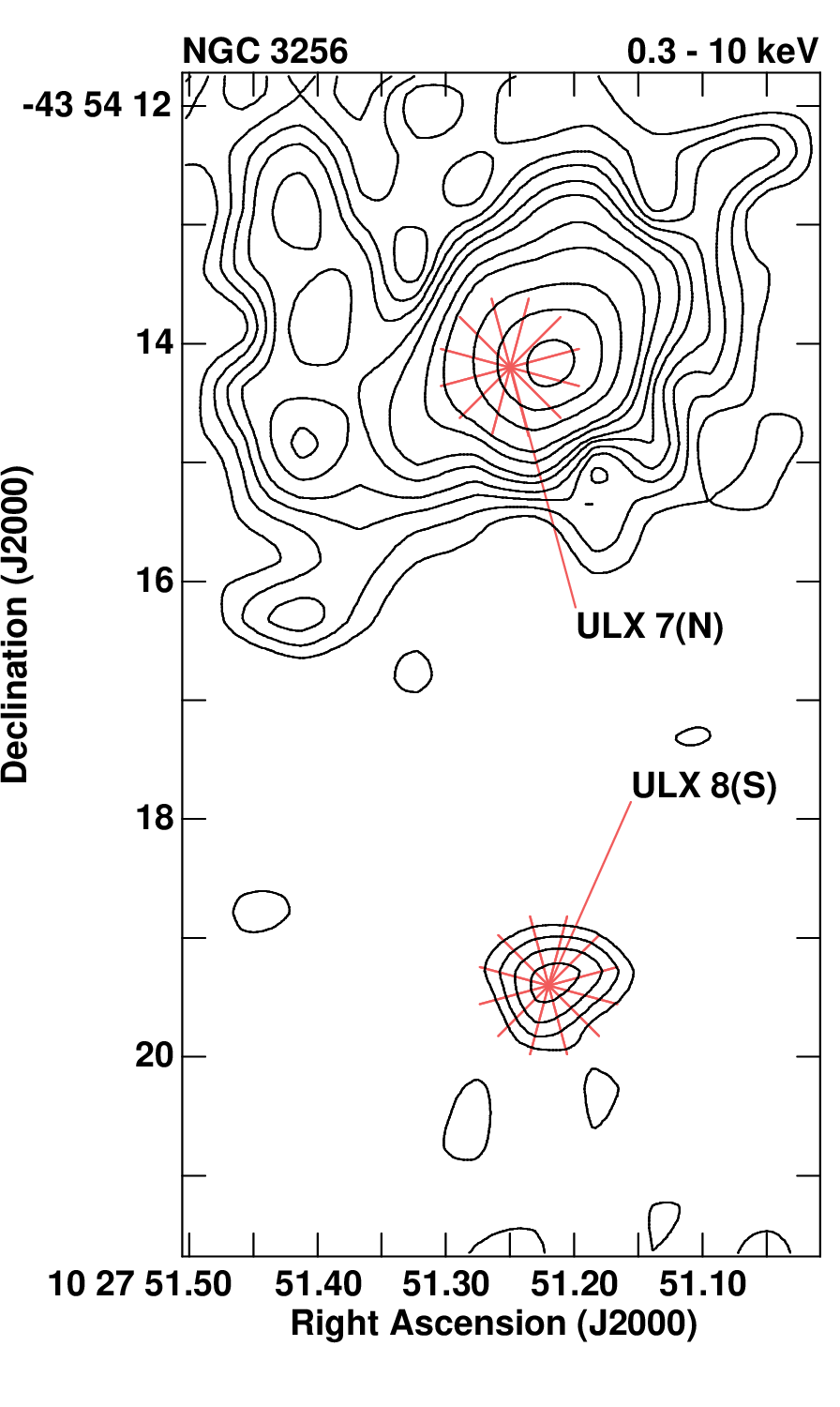}
\caption{Continued}
\label{fig2c}
\end{figure}

\begin{figure}
\epsscale{1.00}
\plotone{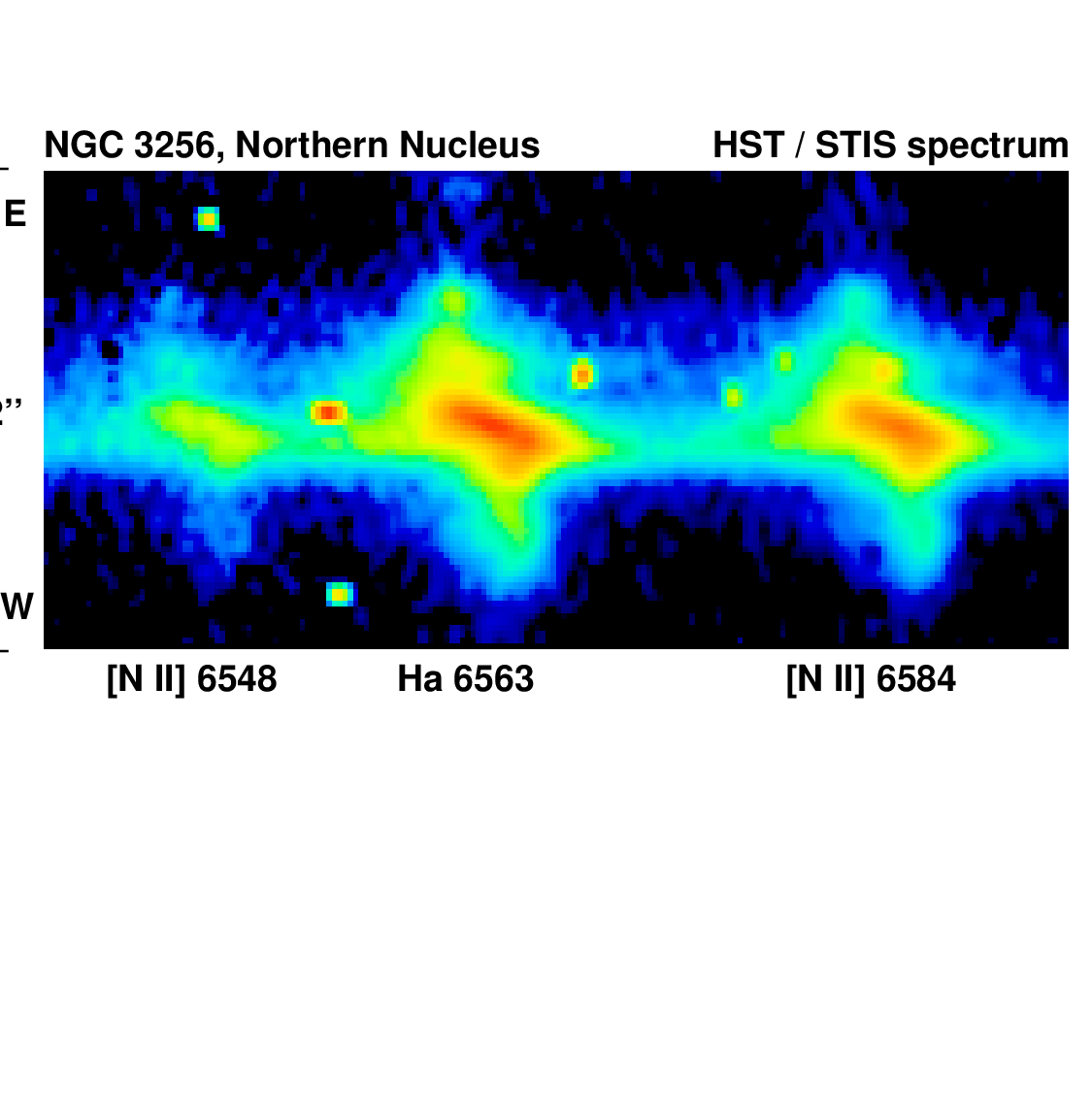}
\caption{ {\it HST} long-slit spectrum of the northern nucleus 
of NGC~3256, showing broad H$\alpha$ and 
[N II]$\lambda\lambda$6548,6584\AA\ emission lines.
The emission lines indicate strong rotation around the
nucleus, with a peak velocity shear 
of $\sim$300km sec$^{-1}$ over a projected distance
of $\sim$40pc. 
The 52\dsec0 $\times$ 0\dsec2 slit was oriented 
at a position angle of $91^\circ$, and two 340sec exposures were
offset slightly along the slit.  These and matching shorter 
wavelength spectra also show broadened lines from
[S II]$\lambda\lambda$6717,6731\AA\ and H$\beta$.  
(There are six blemishes in this subimage due to cosmic 
rays: one at the [N II]$\lambda$6548\AA\ line 
(0\dsec9 above continuum), two between [N II]$\lambda$6548\AA\ 
and H$\alpha$ (one above, one below), one just longward 
of H$\alpha$ and two just shortward of [N II]$\lambda$6584\AA\ 
(all above the continuum).) }
\label{fig3}
\end{figure}


\end{document}